\documentclass{PoS}

\usepackage{amsmath,amssymb}

\def\hc{\text{h.c.}}

\def\G{$\mathcal G$}

\title{Theory status and implications of $R_K^{(\ast)}$}

\ShortTitle{Theory status and implications of $R_K^{(\ast)}$}

\author{Avelino Vicente\\
  Instituto de F\'{\i}sica Corpuscular (CSIC-Universitat de Val\`{e}ncia),
  Apdo. 22085, E-46071 Valencia, Spain\\
  E-mail: \email{avelino.vicente@ific.uv.es}}

\abstract{
The LHCb and Belle collaborations have reported on some anomalies in
$b \to s$ transitions, with discrepancies with the Standard Model
predictions in some angular observables and branching ratios and
intriguing hints for lepton universality violation. We will review the
current situation and explore the proposed New Physics explanations
for these tensions. We will also discuss the possible connection of
the $b \to s$ anomalies to other central problems in physics, such as
the dark matter of the Universe, the origin of neutrino masses or the
strong CP problem.
}

\FullConference{18th International Conference on B-Physics at Frontier Machines - Beauty2019 -\\
		29 September / 4 October, 2019\\
		Ljubljana, Slovenia}

\begin{document}

\section{Introduction}
\label{sec:intro}

Since 2013, B physics has attracted a great deal of attention in the
particle physics community due to a wide set of intriguing
experimental anomalies. In particular, several measurements in
semileptonic processes involving $b \to s$ transitions have been found
to deviate from the Standard Model (SM) predicted values, hence
suggesting the presence of some New Physics (NP) effects. The list of
anomalies includes substantial deviations in the measurements of
branching ratios and angular observables by the
LHCb~\cite{Aaij:2013aln,Aaij:2013qta,Aaij:2015oid,Aaij:2015esa} and
Belle~\cite{Abdesselam:2016llu,Wehle:2016yoi} collaborations, as well
as discrepancies in the observables
\begin{align}
R_{K^{(\ast)}} = \frac{ \Gamma(B \rightarrow K^{(\ast)} \mu^+ \mu^-)}{\Gamma(B \rightarrow K^{(\ast)} e^+ e^-)} \, ,
\end{align}
measured in specific dilepton invariant mass squared ranges $q^2 \in
[q^2_{\rm min}, q^2_{\rm max}]$. The $R_{K^{(\ast)}}$ ratios were
introduced in~\cite{Hiller:2003js} in order to test lepton flavor
universality (LFU), a central feature of the SM that predicts
$R_{K^{(\ast)}} \sim 1$. Very interestingly, the LHCb results for the
$R_K$ ratio in one $q^2$ bin~\cite{Aaij:2014ora,Aaij:2019wad} and the
$R_{K^\ast}$ ratio in two $q^2$ bins~\cite{Aaij:2017vbb} were found to
lie significantly below one:
\begin{align}
R_K &= 0.846^{+0.060}_{-0.054}\text{(stat)}^{+0.016}_{-0.014}\text{(syst)}    \,, \quad
&q^2 \in [1,6]~\text{GeV}^2 \,, \nonumber \\[0.2cm]
R_{K^\ast} &= 0.660^{+0.110}_{-0.070}\text{(stat)}\pm0.024\text{(syst)}    \,, \quad
&q^2 \in [0.045,1.1]~\text{GeV}^2\,, \nonumber \\[0.2cm]
R_{K^\ast} &= 0.685^{+0.113}_{-0.069}\text{(stat)}\pm0.047\text{(syst)}     \,, \quad
&q^2 \in [1.1,6.0]~\text{GeV}^2 \,. 
\end{align}
A comparison between the LHCb results and the SM predictions for $R_K$
and $R_{K^\ast}$ derived
in~\cite{Descotes-Genon:2015uva,Bordone:2016gaq} leads to
discrepancies with the SM above the $2 \, \sigma$ level, with the
precise statistical significance depending on the dilepton invariant
mass squared range considered. Furthermore, it is generally accepted
that unknown QCD effects cannot account for the observed deviations,
since they cancel out to great accuracy in the $R_{K^{(\ast)}}$
ratios. Although these hints are still not very significant, they are
definitely intriguing and, if confirmed, would have dramatic
consequences for the \textit{shape} of NP, which would necessarily
violate LFU.

Many NP models have been proposed in order to explain the $b \to s$
anomalies. The most widely studied scenarios include heavy $Z^\prime$
bosons or leptoquarks, although many other possibilities and
variations of the minimal setups have also been put forward in recent
years. Here we discuss the most popular options and discuss their
possible connection to other open problems in physics. Although we
will not attempt to make a complete review of all the proposed
scenarios, and many well-motivated models will be omitted, we hope
that our discussion constitutes a good description of the panorama
depicted by the model builders.

\section{Interpreting the anomalies}
\label{sec:EFT}

The set of experimental measurements in $b \to s$ transitions can be
interpreted in a model-independent way by using the language of
Effective Field Theory (EFT). This approach is valid if all NP degrees
of freedom have masses well above the energy scale of the observables
of interest, a well-motivated assumption due to the lack of
observations in direct searches, both at the Large Hadron Collider
(LHC) as well as in low-energy experiments. In this case, one can
integrate out all the NP states and describe the physical observables
by a collection of non-renormalizable operators, with canonical
dimensions higher than four.

\begin{figure}[t]
\centering
\includegraphics[width=0.35\textwidth]{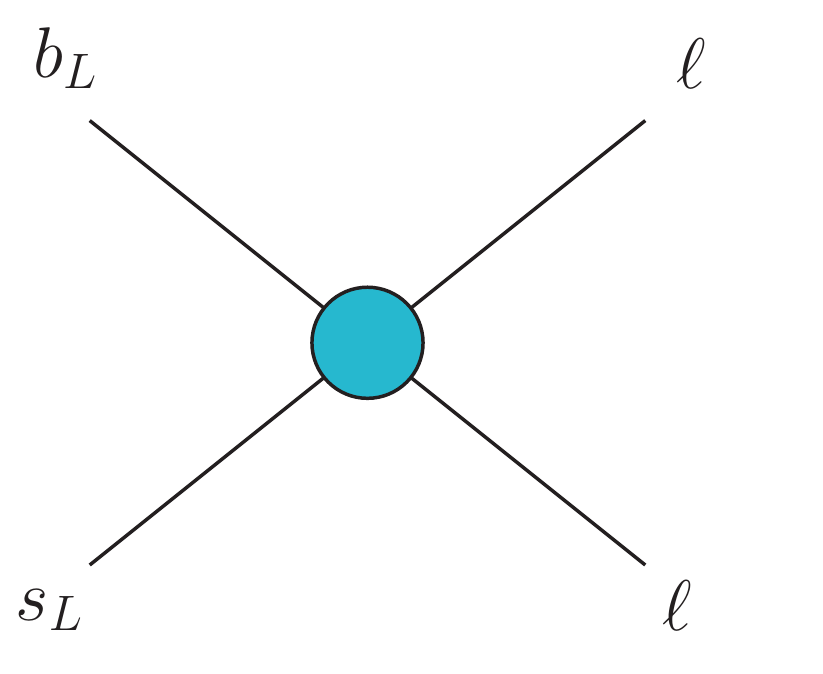}
\caption{$O_9$ operator with muon lepton flavors.}
\label{fig:O9}
\end{figure}

The effective Hamiltonian describing $b \to s$ transitions is typically
written as
\begin{equation} \label{eq:effH}
\mathcal H_{\text{eff}} = - \frac{4 G_F}{\sqrt{2}} \, V_{tb} V_{ts}^\ast \, \frac{e^2}{16 \pi^2} \, \sum_i \left(C_i \, \mathcal O_i + C^\prime_i \, \mathcal O^\prime_i \right) + \hc \, ,
\end{equation}
where $G_F$ denotes the Fermi constant that sets the strength of weak
interactions, $e$ is the electric charge and $V$ the
Cabibbo-Kobayashi-Maskawa (CKM) matrix. $\mathcal O_i$ and $\mathcal
O^\prime_i$ are dimension-6 effective operators contributing to $b \to
s$ quark flavor transitions, while $C_i$ and $C^\prime_i$ are their
Wilson coefficients. Among all possible operators involved in $b \to
s$ semileptonic decays, the following subset turns out to be relevant
for the interpretation of the $b \to s$ anomalies:
\begin{align}
\mathcal O_7 &= \left( \bar s \sigma_{\mu \nu} P_R b \right) \, F^{\mu \nu} \, ,   &   \mathcal O^\prime_7 &= \left( \bar s \sigma_{\mu \nu} P_L b \right) \, F^{\mu \nu} \, , \label{eq:O7} \\
\mathcal O_9 &= \left( \bar s \gamma_\mu P_L b \right) \, \left( \bar \ell \gamma^\mu \ell \right) \, ,   &   \mathcal O^\prime_9 &= \left( \bar s \gamma_\mu P_R b \right) \, \left( \bar \ell \gamma^\mu \ell \right) \, , \label{eq:O9} \\
\mathcal O_{10} &= \left( \bar s \gamma_\mu P_L b \right) \, \left( \bar \ell \gamma^\mu \gamma_5 \ell \right) \, ,   &   \mathcal O^\prime_{10} &= \left( \bar s \gamma_\mu P_R b \right) \, \left( \bar \ell \gamma^\mu \gamma_5 \ell \right) \, . \label{eq:O10}
\end{align}
Here $\ell = e, \mu, \tau$ denotes the lepton flavor. Although the
operators and Wilson coefficients actually have flavor indices, we
will omit them in order to simplify the notation. A diagrammatric
representation of a muonic $O_9$ operator is provided in
Fig.~\ref{fig:O9}. It is common to split the Wilson coefficients in
two pieces: the SM contributions and the NP contributions, defining
\begin{align}
C_7 &= C_7^{\text{SM}} + C_7^{\text{NP}} \, , \\
C_9 &= C_9^{\text{SM}} + C_9^{\text{NP}} \, , \\
C_{10} &= C_{10}^{\text{SM}} + C_{10}^{\text{NP}} \, .
\end{align}
In principle, analogous splittings can be defined for the primed
coefficients, but in this case the SM contributions are rather small
and therefore one simply has $C^\prime_7 \simeq C^{\prime \:
  \text{NP}}_7$, $C^\prime_9 \simeq C^{\prime \: \text{NP}}_9$ and
$C^\prime_{10} \simeq C^{\prime \: \text{NP}}_{10}$. The SM
contributions are calculable~\cite{Descotes-Genon:2013vna}, while the NP contributions remain as free parameters to be determined. The
observables measured by the experimental collaborations can be written
in terms of the $C_i$ and $C^\prime_i$ Wilson coefficients. For
instance, simple expressions for $R_K$ and $R_{K^\ast}$ can be found
in~\cite{Celis:2017doq}. Since the same Wilson coefficients enter
several observables, one expects a pattern of deviations from the SM,
rather than a single anomaly in a specific observable. The strategy to
exploit this fact is therefore clear: the NP contributions can be
determined with a global fit of the observables to experimental data.

Several
groups~\cite{Alguero:2019ptt,Alok:2019ufo,Ciuchini:2019usw,DAmico:2017mtc,Datta:2019zca,Aebischer:2019mlg,Kowalska:2019ley,Arbey:2019duh}
have followed this strategy. The different fits agree qualitatively,
although they differ quantitatively due to differences in the form
factors, the treatment of uncertainties or the computational
techniques used. In all cases, a NP scenario with one or more NP
contributions to the Wilson coefficients is preferred over the pure SM
scenario. In all fits, the muonic $C_9$ coefficient seems to be
crucial. A good fit to data is obtained with a NP contribution to this
coefficient of about 20\% of the SM contribution (and with opposite
sign). Other muonic coefficients may have NP contributions as well and
in fact three competitive 1D (muonic) scenarios emerge:
$C_9^{\text{NP}}$ only, $C_9^{\text{NP}} = - C_{10}^{\text{NP}}$ and
$C_9^{\text{NP}} = -C_9^{\prime \: \text{NP}}$. No indication for NP
contributions in electronic coefficients is found.~\footnote{See also
  the contribution by S\'ebastien Descotes-Genon, also in these
  proceedings.}

These findings provide a quantitive assessment of the $b \to s$
anomalies and serve as a guide for model builders that aim at an
explanation. For instance, valuable information about the scale of NP
can be inferred~\cite{DiLuzio:2017chi}. $C_9^{\text{NP}} \sim - 20\% \, \times \,
C_9^{\text{SM}}$ leads to
\begin{equation}
\frac{C_9^{\rm NP}}{\Lambda_{\rm NP}^2} \,  \sim 20 \% \, \times \, \frac{4 G_F}{\sqrt{2}} \, V_{tb} V_{ts}^\ast \, \frac{e^2}{16 \pi^2} \, C_9^{\rm SM} \, .
\end{equation}
One can then estimate the scale of NP, $\Lambda_{\rm NP}$, in several
generic scenarios:

\begin{itemize}
\item \textit{Unsuppressed NP}: $\displaystyle \quad C_9^{\rm NP} = 1 \quad \Rightarrow \quad \Lambda_{\rm NP} \sim 30$ TeV. 
\item \textit{CKM-suppressed NP}: $\displaystyle \quad C_9^{\rm NP} = |V_{tb} V_{ts}^\ast| \quad \Rightarrow \quad \Lambda_{\rm NP} \sim 6$ TeV. 
\item \textit{Loop-suppressed NP}: $\displaystyle \quad C_9^{\rm NP} = \frac{1}{16 \pi^2} \quad \Rightarrow \quad \Lambda_{\rm NP} \sim 2.50$ TeV. 
\item \textit{CKM\&loop-suppressed NP}: $\displaystyle \quad C_9^{\rm NP} = \frac{|V_{tb} V_{ts}^\ast|}{16 \pi^2} \quad \Rightarrow \quad \Lambda_{\rm NP} \sim 0.5$ TeV. 
\end{itemize}

This way, one concludes that only when the NP effects are suppressed
(by CKM factors and/or loops), the scale of NP is low enough to be
probed directly at current facilities. In what concerns the NP
mediators behind the $O_i$ operators, only two possibilities exist at
tree-level: a neutral $Z^\prime$ vector boson and a scalar or vector
leptoquark. In fact, these have been the most popular options in the
literature. Other possibilities at loop level have also been
considered. We now proceed to discuss several example models aiming at
an explanation to the $b \to s$ anomalies.

\section{Models for $R_K^{(*)}$}
\label{sec:models}

The $b \to s$ anomalies have triggered the creativity of model
builders and the construction of many new data-driven models. Some of
these models would have never been built in the absence of the strong
motivation emerged from the intriguing experimental hints in B-meson
decays, which have opened new directions beyond the SM. Therefore,
even if the anomalies go away, the model building community has found
novel ways to address some of the most important problems in the SM
and shed light on some of the crucial questions, such as the flavor
problem, the dark matter of the Universe or the origin of neutrino
masses.

We will now review some of the most popular setups proposed to explain
the $b \to s$ anomalies.

\subsection{$Z^\prime$ models}
\label{subsec:Zprime}

\begin{figure}[t]
\centering
\includegraphics[width=0.45\textwidth]{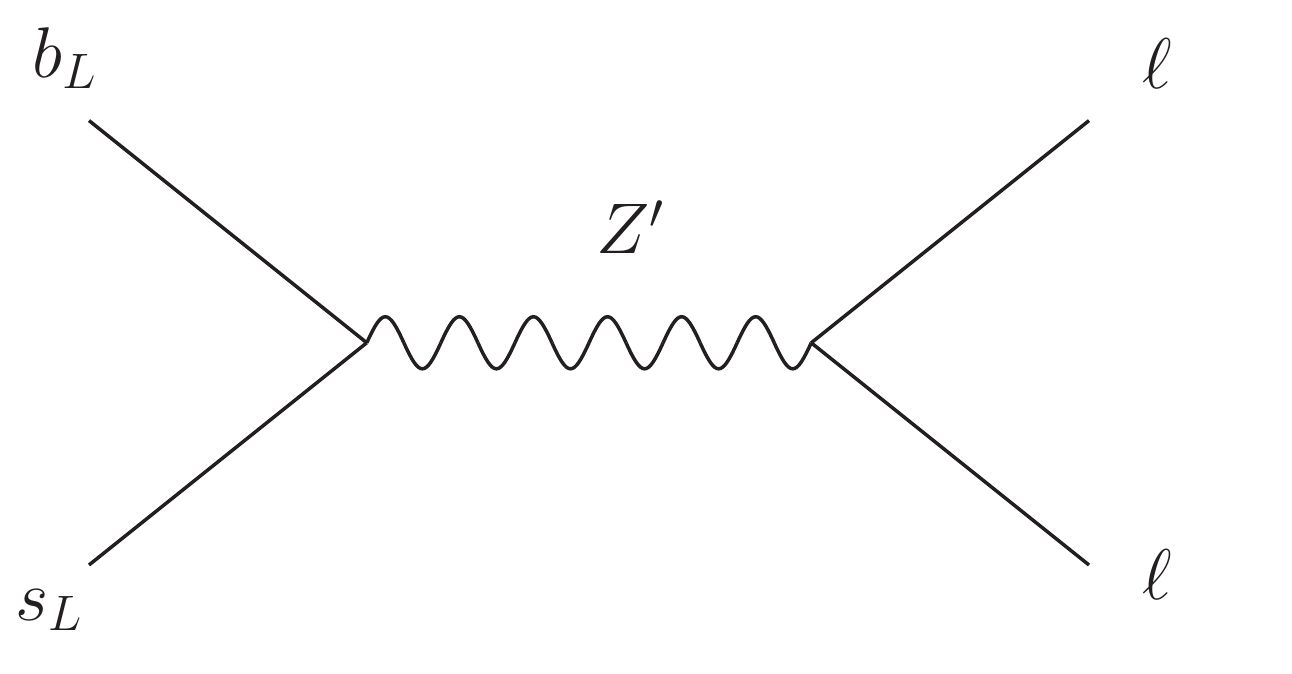}
\caption{Generation of the $O_9$ operator due to the exchange of a
  neutral $Z^\prime$ vector boson.}
\label{fig:Zprime}
\end{figure}

$Z^\prime$ models argueably constitute the easiest solution to the $b
\to s$ anomalies. A neutral $Z^\prime$ boson can generate the $O_9$
operator (and perhaps $O_{10}$ or $O_9^\prime$ as well) as shown in
Fig.~\ref{fig:Zprime}. The list of requirements for the model is
rather short: a $Z^\prime$ boson that contributes to the desired
operators, with flavor violating couplings to quarks and non-universal
couplings to leptons.~\footnote{Optionally, but highly desirable, is
  to have a rich interplay with some other physics, as discussed in
  Sec.~\ref{sec:connection}.} Such vector boson would obtain its mass
from the spontaneous breaking of an enlarged gauge symmetry group \G,
which includes the SM gauge group $\mathcal G_{\rm SM} = \rm{SU(3)_c
  \times SU(2)_L \times U(1)_Y}$ as a subgroup and gets broken at the
high-energy scale $\Lambda_{\rm NP} \sim m_{Z^\prime}$. The simplest
solution is to add a new $\rm U(1)$ factor to the SM gauge group, but
other more involved possibilities also exist.~\footnote{See for
  instance \cite{Boucenna:2016wpr,Boucenna:2016qad} for a $Z^\prime$
  model for the $b \to s$ anomalies based on the addition of a
  non-Abelian $\rm SU(2)$ gauge group.} Regarding the non-universal
couplings to the SM charged leptons, one can broadly classify all
$Z^\prime$ models in two categories:

\begin{itemize}
\item \textit{Direct models}: in these models the $Z^\prime$ boson
  couples directly to the SM charged leptons, and therefore these must
  be charged non-universally under \G.
\item \textit{Indirect models}: in these models the SM charged leptons
  are universally charged under \G, and the non-universality is
  induced by their mixing with new fermionic states which have a
  different representation under \G.
\end{itemize}

{
\renewcommand{\arraystretch}{1.4}
\begin{table}
\centering
\begin{tabular}{ccccc} 
\hline \hline 
Field & Spin  & \( \rm SU(3)_c \times\, SU(2)_L \times\, U(1)_Y \times U(1)_X \) \\ 
\hline
\(\phi\) & \(0\)  & \(({\bf 1}, {\bf 1}, 0, 2) \) \\
\(Q_{L,R}\) & \(\frac{1}{2}\)  & \(({\bf 3}, {\bf 2}, \frac{1}{6}, 2) \) \\ 
\(L_{L,R}\) & \(\frac{1}{2}\)  & \(({\bf 1}, {\bf 2}, -\frac{1}{2}, 2) \) \\ 
\hline \hline
\end{tabular} 
\caption{New scalars and fermions in the model of \cite{Sierra:2015fma}.}
\label{tab:DarkBS}
\end{table}
}

Let us illustrate the category of indirect models with the setup
introduced in~\cite{Sierra:2015fma}. The model extends $\mathcal
G_{\rm SM}$ with a new $\rm U(1)_X$ factor, under which all the SM
particles are singlets. The particle content of the model includes the
new scalar $\phi$ as well as the new vector-like fermions $Q$ and $L$,
charged under $\rm U(1)_X$.~\footnote{Actually, the original version
  of the model also includes the scalar $\chi$, which serves as a dark
  matter candidate. This \textit{optional} feature is not relevant for
  the current discussion and will be mentioned in
  Sec.~\ref{subsec:dm}. Also, we note that a variation of this model
  including non-zero neutrino masses was discussed
  in~\cite{Rocha-Moran:2018jzu}.} We note that $\phi$ is a singlet of
$\mathcal G_{\rm SM}$, while $Q$ and $L$ have the same representation
under $\mathcal G_{\rm SM}$ as the SM quark and lepton doublets $q$
and $\ell$. Due to their vector-like nature, one can write
gauge-invariant Dirac mass terms for $Q$ and $L$,
\begin{equation} \label{eq:VectorMass}
\mathcal L_m = m_Q \, \overline Q Q + m_L \, \overline L L \, .
\end{equation}
Furthermore, they also have Yukawa interactions
\begin{equation} \label{eq:VectorYukawa}
\mathcal L_Y = \lambda_Q \, \overline{Q_R} \, \phi \, q_L + \lambda_L \, \overline{L_R} \, \phi \, \ell_L + \hc \, .
\end{equation}
Here $\lambda_Q$ and $\lambda_L$ are $3$-component Yukawa vectors. In what concerns the scalar sector, we will simply assume that the SM gauge symmetry is broken as usual and that $\phi$ gets the vacuum expectation value
\begin{equation}
\langle \phi \rangle = \frac{v_\phi}{\sqrt{2}} \, ,
\end{equation}
hence breaking the new $\rm U(1)_X$ piece and leading to a massive
$Z^\prime$ boson, with mass $m_{Z^\prime} = 2 g_X v_\phi$, with $g_X$
the $\rm U(1)_X$ gauge coupling.

\begin{figure}
\centering
\includegraphics[scale=0.5]{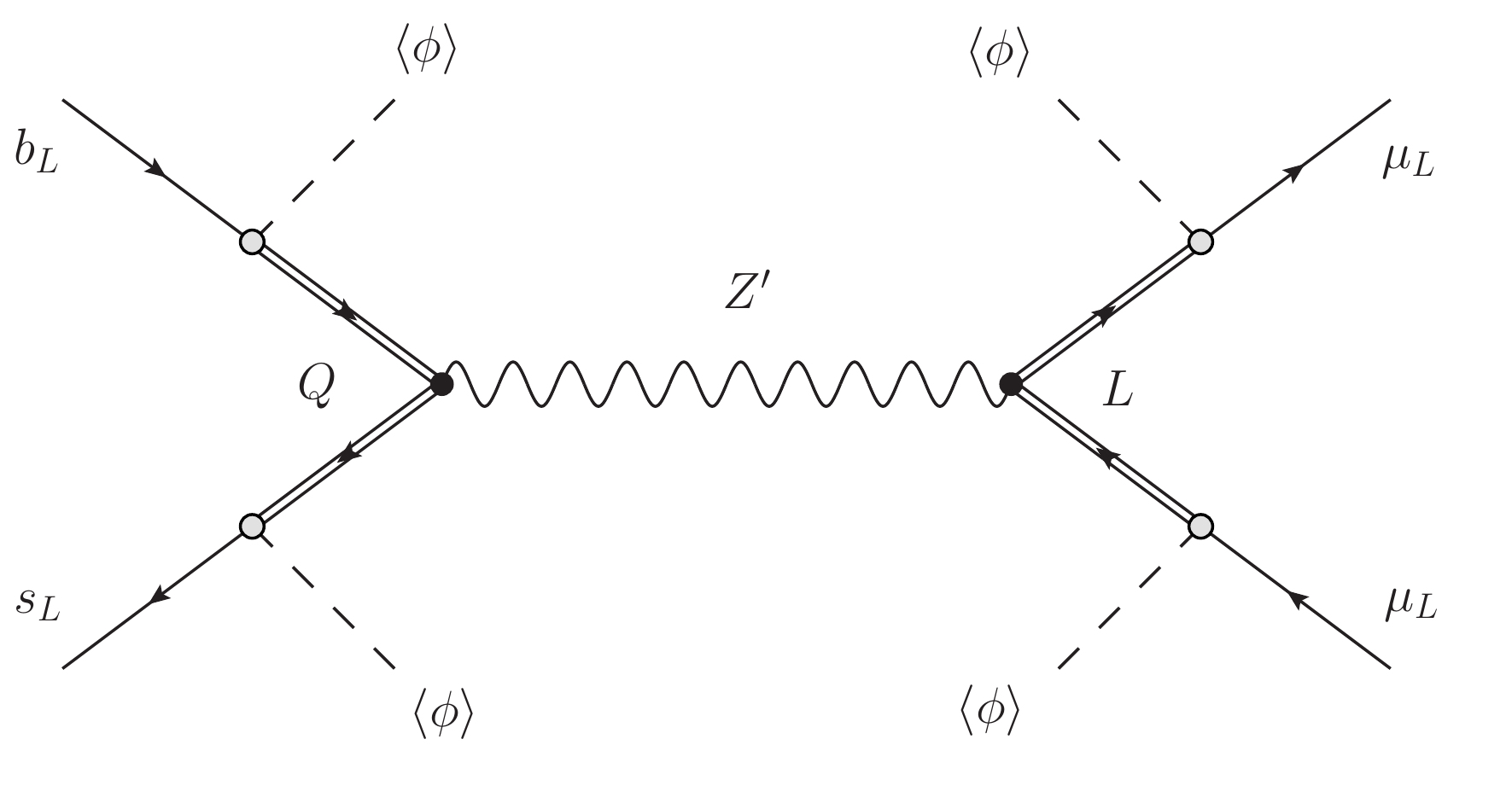}
\caption{Generation of $O_9$ and $O_{10}$ in the model of
  \cite{Sierra:2015fma}. Since the $Z^\prime$ couplings to fermions
  are purely left-handed, this model predicts $C_9^{\rm NP} = -
  C_{10}^{\rm NP}$.}
\label{fig:couplings}
\end{figure}

Let us now come to the $b \to s$ anomalies and their resolution in
this model. One should note that in this model the SM fermions do not
couple \textit{directly} to the $Z^\prime$ boson. These couplings are
only generated after symmetry breaking due to the mixing with the
vector-like fermions. This effectively leads to the generation of
$O_9$ and $O_{10}$, as shown in Fig.~\ref{fig:couplings}. The
non-universality in the charged lepton couplings to the $Z^\prime$
boson originates simply from the non-universality of the $\lambda_L$
Yukawa couplings. For instance, assuming $\lambda_L^e = \lambda_L^\tau
= 0$, the $Z^\prime$ boson couples to muons only, easily explaining
the anomalies in $R_K$ and $R_{K^\ast}$.

Many \textit{direct models} also exist, in some cases in combination
with \textit{indirect} couplings in the quark or lepton sectors.  A
very popular model is the one based on the $L_\mu - L_\tau$ gauge
symmetry, pioneered in the context of the $b \to s$ anomalies in the
relevant paper~\cite{Altmannshofer:2014cfa} and later explored in many
other works. An extension of the $L_\mu - L_\tau$ gauge group to
include the quark sector was considered
in~\cite{Crivellin:2015lwa}. Other possibilities have also been
discussed in the literature. For instance, the gauged BGL
symmetry~\cite{Branco:1996bq} introduced in~\cite{Celis:2015ara} and
the $\rm U(2)$ flavor symmetry in~\cite{Falkowski:2015zwa}, to mention
two particularly attractive proposals.

A generally very relevant constraint in all $Z^\prime$ models for the
$b \to s$ anomalies is $B_s - \overline{B_s}$ mixing, recently
discussed in detail in~\cite{DiLuzio:2019jyq}. The potential conflict
originates from the fact that this mixing is induced at tree-level and
depends on the same $Z^\prime - b - s$ coupling required to explain
the anomalies. In models with both left- and right-handed $Z^\prime$
couplings to quarks, one can in principle get a cancellation that
alleviates the tension~\cite{Crivellin:2015era}, but in models with
only one chirality this is not possible~\cite{Rocha-Moran:2018jzu}. In
such cases, the usual solution is to suppress the $Z^\prime - b - s$
coupling at the cost of increasing the $Z^\prime - \mu - \mu$
coupling, so that the NP contributions to the muonic $C_9$
coefficients are large enough to explain the anomalies.

\subsection{Leptoquark models}
\label{subsec:LQ}

\begin{figure}[t]
\centering
\includegraphics[width=0.35\textwidth]{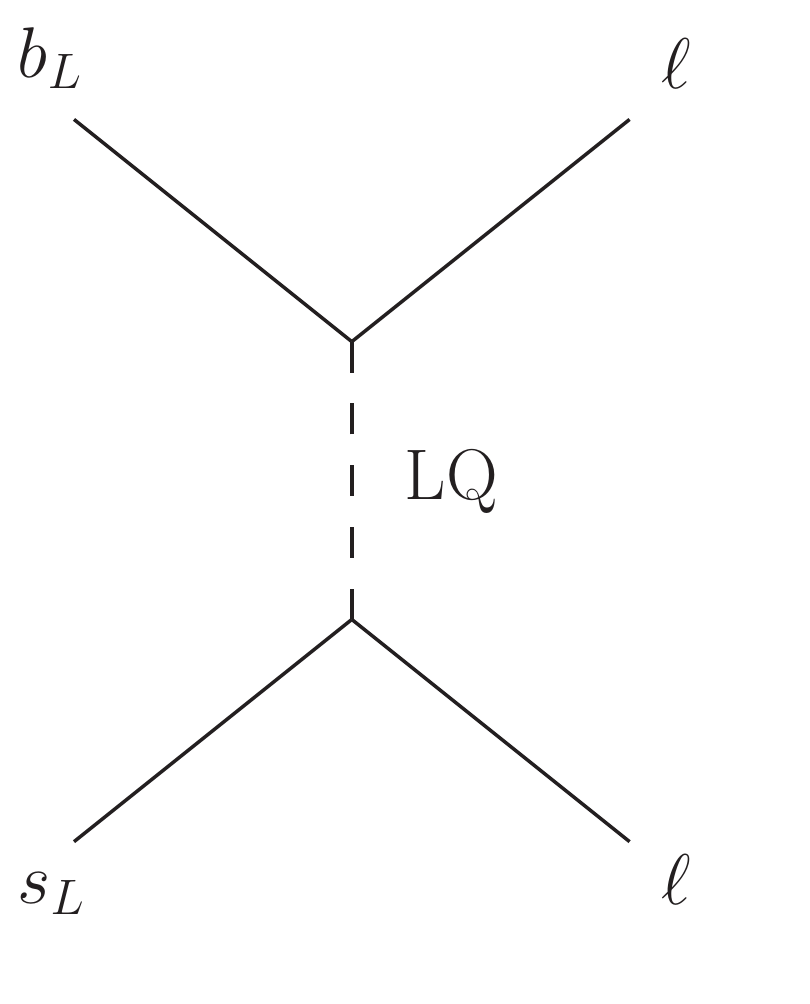}
\caption{Generation of the $O_9$ operator due to the exchange of a leptoquark. The leptoquark can be either scalar or vector.}
\label{fig:LQ}
\end{figure}

Leptoquarks~\cite{Dorsner:2016wpm} are scalars or vectors that couple
simultaneously to both quarks and leptons. While they appear naturally
in grand unified theories, there is in principle no fundamental reason
why they cannot have masses well below the unification scale and have
sizable effects in flavor observables. In fact, leptoquark models are
also very popular and many simple solutions to the $b \to s$ anomalies
have been proposed.~\footnote{Again, it is highly desirable to be able
  to address other physics problems, and several examples linking
  leptoquarks, $b \to s$ anomalies and other open questions are
  discussed in Sec.~\ref{sec:connection}.} Fig.~\ref{fig:LQ} shows how
to generate the $O_9$ operator via the tree-level exchange of a
leptoquark.

Leptoquarks have been widely discussed recently due to their potential
to explain the experimental anomalies observed not only in neutral $b
\to s$ transitions, the subject here, but also in charged $b \to c$
decays. We refer to other contributions to these conference
proceedings where the phenomenology of leptoquarks and their role in
solving the $b \to c$ anomalies are discussed in great
detail.~\footnote{See contributions by Nejc Ko\v{s}nik and Teppei
  Kitahara, also in these proceedings.}

\subsection{Loop models}
\label{subsec:loop}

Finally, loop models explain the $b \to s$ anomalies by generating the
$O_9$ operator at loop level. Although this allows for a wider variety
of options, since many possibilities for the particles running in the
loop exist, fewer models with this feature have been built. The
different one-loop models that one can construct in connection to the
$b \to s$ anomalies have been classified
in~\cite{Gripaios:2015gra,Arnan:2016cpy}. Interestingly, the
constraints from $B_s - \overline{B_s}$ mixing, so important in case
of $Z^\prime$ models, can be very different depending on the charges
chosen for the new states running in the loop. Also, as we will
discuss below, some of the possible loop models include particles with
the correct quantum numbers to be valid dark matter candidates, a
possibility that has been explored in some works.

\section{Connection to other physics problems}
\label{sec:connection}

All models addressing the $b \to s$ anomalies include additional
fields and, in some cases, new gauge symmetries. This offers many new
phenomenological opportunities, as one would expect that the new
states communicate not only to the sector strictly related to the
anomalies, but also to other SM sectors. The natural question is then:
what if the explanation to the $b \to s$ anomalies also solves other
physics problems? In this Section we briefly discuss some of the ideas
that have been presented in this direction, linking the $b \to s$
anomalies to dark matter, the origin of neutrino masses and the strong
CP problem.

\subsection{Dark matter}
\label{subsec:dm}

The possible link between the $b \to s$ anomalies and the dark matter
of the Universe was reviewed in~\cite{Vicente:2018xbv}. Here we will
briefly discuss two representative examples.

Let us first consider the addition of the complex scalar $\chi$ to the
model discussed in Sec.~\ref{subsec:Zprime}, with representation
$({\bf 1}, {\bf 1}, 0, -1)$ under the extended gauge group. This was
actually the original version of the model introduced
in~\cite{Sierra:2015fma}. Assuming that $\chi$ does not get a vacuum
expectation value, the spontaneous breaking of the $\rm U(1)_X$
symmetry leaves a remnant $\mathbb{Z}_2$ \textit{dark parity}, under
which $\chi$ is odd and the rest of states are even. Therefore, $\chi$
is perfectly stable, without any additional ad-hoc symmetry. The same
gauge symmetry behind the dynamics required to explain the $b \to s$
anomalies stabilizes the dark matter candidate.~\footnote{The
  generation of remnant $\mathbb{Z}_N$ symmetries from continuous $\rm
  U(1)$ groups has been discussed
  in~\cite{Krauss:1988zc,Petersen:2009ip,Sierra:2014kua}.}
Furthermore, $\chi$ is a singlet under the SM gauge group and does not
have any Yukawa interaction. Assuming that the Higgs portal $|H|^2
|\chi|^2$ term is suppressed, it can only be produced in the early
Universe via $Z^\prime$ exchange. This establishes another link to the
$b \to s$ anomalies. Ref.~\cite{Sierra:2015fma} shows that one can
explain the anomalies in a region of the parameter space that can also
reproduce the observed dark matter relic density.
 
A second example was provided in~\cite{Kawamura:2017ecz}. In this case
a loop model for the $b \to s$ anomalies was considered. The particle
content and gauge charges of the model is the same as in the model
discussed in Sec.~\ref{subsec:Zprime}, but with a crucial difference:
the $\rm U(1)_X$ symmetry is assumed to be global and
conserved. Therefore, the $b \to s$ anomalies cannot be explained via
$Z^\prime$ exchange, but the $O_9$ and $O_{10}$ operators are
generated at the one-loop level, also with $C_9^{\rm NP} = -
C_{10}^{\rm NP}$. Regarding dark matter, the conservation of $\rm
U(1)_X$ stabilizes the lightest particle charged under this symmetry,
and this is taken to be the scalar $\phi$. Again, the authors
of~\cite{Kawamura:2017ecz} explicitly show that one can simultaneously
explain the $b \to s$ anomalies and reproduce the measured dark matter
relic density. Finally, the model is found to be testable in future
direct DM detection experiments as well as by direct LHC searches.

\subsection{Neutrino masses}
\label{subsec:nu}

The main open question in the lepton sector is the origin of neutrino
masses. What if the $R_K$ and $R_{K^\ast}$ LFU violating hints
(remember: L stands for \textit{lepton}!) can guide us towards solving
this central problem?  While this seems like a very natural question,
it has been explored very little in recent years.

Leptoquarks are \textit{familiar} states in the neutrino mass
model-building community. It is well known that the addition of two
leptoquarks (or a leptoquark and another exotic state) allows one to
induce radiative Majorana neutrino masses~\cite{Cai:2017jrq}, and this
mechanism has been considered in several leptoquark models for the
B-anomalies~\cite{Pas:2015hca,Guo:2017gxp,Hati:2018fzc}. Furthermore,
the possible link between the violation of LFU and mixing in the
leptonic sector has been discussed in
\cite{Boucenna:2015raa,Botella:2017caf}.

\subsection{Strong CP problem}
\label{subsec:cp}

Many of the models capable to address all the experimental anomalies
in B-meson decays, both in neutral $b \to s$ and in charged $b \to c$
transitions, include the $U_1 \sim ({\bf 3}, {\bf 1}, 2/3)$ vector
leptoquark.~\footnote{See~\cite{Buttazzo:2017ixm} for a review of
  combined explanation to all B-anomalies.} This state has been shown
to provide a good simultaneous explanation, but it also requires to
extend the SM gauge group in order to embed $U_1$ as one of the gauge
bosons. This has led to the so-called 4321 models, based on the
extended gauge group $\mathcal{G}_{4321}\equiv \rm SU(4)\times
SU(3)^\prime\times SU(2)_L\times
U(1)^\prime$~\cite{DiLuzio:2017vat,Bordone:2017bld}. A remarkable
feature of these models is that QCD emerges at low energies from the
product of two non-Abelian gauge groups: $\mathrm{SU(3)_c} = \left[
  \mathrm{SU(3)_4\times SU(3)^\prime} \right]_{\rm diag}$, where $\rm
SU(3)_4$ is a subgroup of $\rm SU(4)$. As shown
in~\cite{Fuentes-Martin:2019bue}, the resolution of the strong CP
problem \textit{\`a la Peccei-Quinn} requires in this case the
introduction of two axions. While the properties of the lightest one
are those of the standard QCD axion, a second heavier axion must
exist, with its mass and couplings closely related to the B-anomalies.

\section{Summary}

We are living exciting times in the flavor community, with the
anomalies in B-meson decays providing a strong motivation for novel
research in various directions. Here we have reviewed the current
situation and discussed the different types of models proposed to
address the anomalies in $b \to s$ transitions. We have also pointed
out that $R_K$ and $R_{K^\ast}$ might just be the tip of the iceberg,
with a whole NP sector close to be found, with potential impact in
other central physics problems, such as the dark matter of the
Universe, the origin of neutrino masses or the strong CP problem.

\section*{Acknowledgements}

This work has been supported by the Spanish grants
FPA2017-85216-P~(MINECO/AEI/FEDER, UE), SEJI/2018/033~(Generalitat
Valenciana) and FPA2017-90566-REDC~(Red Consolider MultiDark). The
author acknowledges financial support from MINECO through the Ram\'on
y Cajal contract RYC2018-025795-I.

\end{document}